\documentclass[aps,twocolumn,pra,showpacs,tightenlines]{revtex4-1}
\usepackage{amsmath}
\usepackage{graphicx}
\usepackage{color}
\usepackage{amsfonts}
\usepackage{txfonts}
\usepackage[colorlinks,citecolor=blue]{hyperref}

\begin{document}
\title{Generation of macroscopic entangled cat states in a molecular cavity-QED system}
\author{Jian Huang}
\affiliation{Key Laboratory of Low-Dimensional Quantum Structures and Quantum Control of Ministry of Education, Department of Physics and Synergetic Innovation Center for Quantum Effects and Applications, Hunan Normal University, Changsha 410081, China}
\author{Yu-Hong Liu}
\affiliation{Key Laboratory of Low-Dimensional Quantum Structures and Quantum Control of Ministry of Education, Department of Physics and Synergetic Innovation Center for Quantum Effects and Applications, Hunan Normal University, Changsha 410081, China}
\author{Jin-Feng Huang}
\email{jfhuang@hunnu.edu.cn}
\affiliation{Key Laboratory of Low-Dimensional Quantum Structures and Quantum Control of Ministry of Education, Department of Physics and Synergetic Innovation Center for Quantum Effects and Applications, Hunan Normal University, Changsha 410081, China}
\author{Jie-Qiao Liao}
\email{jqliao@hunnu.edu.cn}
\affiliation{Key Laboratory of Low-Dimensional Quantum Structures and Quantum Control of Ministry of Education, Department of Physics and Synergetic Innovation Center for Quantum Effects and Applications, Hunan Normal University, Changsha 410081, China}

\begin{abstract}
Macroscopic entangled cat states not only are significant in the demonstration of the fundamentals of quantum physics, but also have wide applications in modern quantum technologies such as continuous-variable quantum information processing and quantum metrology. Here we propose a scheme for generation of macroscopic entangled cat states in a molecular cavity-QED system, which is composed of an organic molecule (including electronic and vibrational states) coupled to a single-mode cavity field. By simultaneously modulating the resonance frequencies of the molecular vibration and the cavity field, the molecular vibrational displacement can be enhanced significantly and hence macroscopic entangled cat states between the molecular vibrational mode and the cavity mode can be created. We also study quantum coherence effects in the generated states by calculating the joint Wigner function and the degree of entanglement. The dissipation effects are included by considering the state generation in the open-system case. Our results will pave the way to the study of quantum physics and quantum chemistry in molecular cavity-QED systems.
\end{abstract}

\maketitle

\section{Introduction\label{sec:intro}}

Molecular cavity quantum electrodynamics (QED) is an emerging research field addressing the interactions between the cavity fields and the molecular states~\cite{Haroche1989,Berman1994,WangD2019,Flick2017,Shalabneya2015,Pino2015}. According to different coupling forms, there exist several typical molecular cavity-QED systems~\cite{Roelli2015,Schm2016,Neuman2019,Pino2016,Wu2016,Galego2015}. One typical system considers a full quantum description of the interaction between the plasma and the molecular vibration~\cite{Roelli2015,Schm2016,Neuman2019}, while this model fails to describe the rich structure provided by the internal electronic degree of freedom. At the same time, some works have proposed another typical molecular cavity-QED system: the organic molecules including electronic and vibrational states are collectively coupled to the confined electromagnetic modes~\cite{Pino2016,Wu2016,Galego2015}, which is known as organic microcavities having very large optical nonlinearities and a wide range of applications in nonlinear spectrum~\cite{Pinoa2015}, control of chemistry~\cite{Galego2016,Felipe2016}, and vibronic polaritons~\cite{Strashko2016,Felipe2017,Ahsan2018}. The prominent properties of this system are that these bosonic modes (the molecular vibration and the electromagnetic field) can be strongly coupled to the electronic states and they are convenient to be regulated via laser driving. In particular, recent experiments have achieved the strong-coupling regime of single molecule at room temperature~\cite{Benz2016,Chikkaraddy2016}. In addition, the molecular cavity-QED systems have been proposed to study the cavity optomechanics in the molecular systems~\cite{Roelli2015,Schm2016,Neuman2019}.

The diverse interactions among these degrees of freedom in molecular cavity-QED systems have wide applications in the frontiers of quantum physics~\cite{WangD2017,Long2018} and quantum chemistry~\cite{Galego2016,Lombardi2018,Lacombe2019}. As an example, one possible application associated with the conditional displacement interactions in molecular cavity-QED systems is to create superposed coherent states~\cite{Milburn1985,Milburn1986,Yurke1986,Armour2002,Marshall2003,Liao2016,Liao2016b} and entangled coherent states~\cite{Sanders1992,Sanders2012}, which can be widely used in modern quantum technologies such as quantum metrology~\cite{Joo2011,Joo2012} and quantum teleportation based on coherent-state bases~\cite{Wang2001,vanEnk2001,Johnson2002}. In these systems, the molecular vibration and the electromagnetic fields can provide natural candidates for local information memory and information transfer carrier in continuous-variable quantum information processing, respectively~\cite{Braunstein2005}. Usually, these superposed coherent-state components should be 
distinguishable with each other such that these states can be coded as quasi-qubit states. Therefore, how to generate macroscopic entangled cat states with distinguishable superposition components becomes an important task. So far, many proposals for generation of the entangled coherent states have been suggested in many physical systems, including ion traps~\cite{Gerry1997,Munro2001}, cavity-QED systems~\cite{Kim1999,Solano2003,Akram2013}, and atomic-BEC systems~\cite{Kuang2003}. However, it remains a great challenge to generate entangled cat states with macroscopically distinct coherent-state components in typical bosonic systems~\cite{Liao2016,Liao2016b}. In current experimental conditions, the amplitude of the conditionally generated coherent states is limited because the displacement coupling strength is much smaller than the resonance frequency of the bosonic mode, and hence it is hard to create entangled and superposed coherent states with distinguishable superposition components.

In this paper, we propose a scheme to generate macroscopic entangled cat states in a molecular cavity-QED system involving the couplings of a single-mode cavity field with an organic molecule. This is achieved by introducing the simultaneous resonance-frequency modulation to the molecular vibration and the cavity field. The physical principle behind the amplitude amplification for the coherent states is that the Floquet sideband modes compensate the driving detuning between the molecular vibration and the electronic state displacement force in the electronic-vibrational interaction term. When the effective detuning is tuned to be much smaller than the conditional displacement coupling strength, the displacements of the cavity field and the molecular vibration can be enhanced significantly to be larger than the zero-point motions. This scheme will not only provide a theoretical method for the preparation of macroscopically distinguishable entangled cat states in molecular cavity-QED systems, but also pave the way to the study of the fundamentals of quantum physics and the applications of quantum effects in these systems.

The rest of this paper is organized as follows. In Sec.~\ref{sec:model}, we will introduce the physical model and its Hamiltonian. An approximate Hamiltonian will be derived to depict the evolution of the system. In Sec.~\ref{sec:stategeneration}, we will study the creation of macroscopic entangled cat states with the approximate Hamiltonian and investigate the nonclassical properties of the generated states. Moreover, we will calculate numerically the exact state generation and verify the validity of the used rotating-wave approximation. In Sec.~\ref{sec:opensys}, we will present numerical simulations of the state generation in the open-system case and study the influence of the dissipations on the state generation.  Finally, we will present some discussions in Sec.~\ref{sec:discussion} and conclude this work in Sec.~\ref{sec:conclusion}, respectively. We also present an Appendix to show the derivation of the unitary evolution operator associated with the approximate Hamiltonian.

\section{Model and Hamiltonian \label{sec:model}}
\begin{figure}[tbp]
\center
\includegraphics[bb=-10 0 494 340, width=0.48 \textwidth]{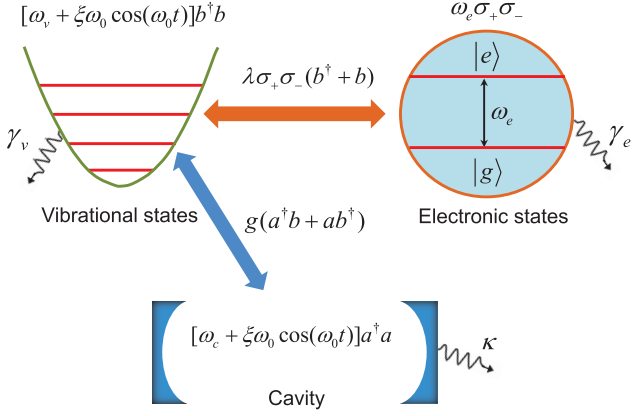}
\caption{Schematic diagram of the molecular cavity-QED system, which is composed of an organic molecule (including the electronic and vibrational states) coupled to a single-mode cavity field. The electronic states with an energy separation $\omega_{e}$ are coupled to the vibrational mode with a resonance frequency $\omega_{v}$ via a conditional displacement interaction $\lambda\sigma_{+}\sigma_{-}(b^{\dag}+b)$, the cavity field is coupled to the vibrational mode via an excitation-exchange-type interaction $g(a^{\dag}b+b^{\dag}a)$. The monochromatic resonance-frequency nodulations are introduced to the molecular vibrational mode and the cavity mode, with the modulation frequency $\omega_{0}$ and modulation magnitude $\xi\omega_{0}$. The decay rates of the electronic state, the vibrational mode, and the cavity mode are denoted by $\gamma_{e}$, $\gamma_{v}$, and $\kappa$, respectively.}
\label{Fig1}
\end{figure}

We consider a molecular cavity-QED system which is composed of an organic molecule coupled to a single-mode cavity field. Here, both the electronic and vibrational degrees of freedom of the molecule are considered, and the coupling between the electronic and vibrational states is described by the Holstein Hamiltonian (with $\hbar=1$)~\cite{Holstein1959,Spano2010,Reitz2019}
\begin{equation}
H_{e-v}=\omega_{e}\sigma_{+}\sigma_{-}+\omega_{v}b^{\dag}b+\lambda\sigma_{+}\sigma_{-}(b^{\dag}+b),\label{Hamilthh}
\end{equation}
where $b$ $(b^{\dagger})$ is the annihilation (creation) operator of the optically active vibrational mode of the molecule with the resonance frequency $\omega_{v}$. Meanwhile, the electronic degree of freedom of the molecule is described by a two-level system with excited state $|e\rangle$ and ground state $|g\rangle$. The energy separation between the two states $|e\rangle$ and $|g\rangle$ is $\omega_{e}$, and the raising and lowering operators of the two-level system are defined by $\sigma_{\pm}$=$(\sigma_{x}\pm i\sigma_{y})/2$, based on the Pauli operators $\sigma_{x}=\vert e\rangle\langle g\vert+\vert g\rangle\langle e\vert $, $\sigma _{y}=i(\vert g\rangle\langle e\vert-\vert e\rangle\langle g\vert) $, and $\sigma_{z}=\vert e\rangle\langle e\vert-\vert g\rangle\langle g\vert$. The interaction between the electronic and vibrational states is characterized by the Huang-Rhys parameter $\lambda$~\cite{Spano2010}, which quantifies the phonon displacement between the ground and excited electronic states.

The molecule is placed inside a cavity and we consider the coupling between the vibrational mode and the cavity field rather than the electronic-state-cavity-field interaction. This is because the strength of the former coupling is much stronger than that of the latter coupling~\cite{Strashko2016}. In particular, the resonant coupling between the cavity mode and the vibrational mode can be achieved by engineering the frequency $\omega_{c}$ of the cavity field such that the cavity mode is resonant or near resonant to the vibrational mode of a specific bond in the molecule~\cite{Shalabneyb2015}. Then the Hamiltonian of the cavity field and its coupling to the vibrational state of the molecule can be written as~\cite{Pinoa2015,Strashko2016}
\begin{equation}
H_{c-v}=\omega_{c}a^{\dag}a+g(a^{\dag}b+ab^{\dag}),\label{Hamilthc}
\end{equation}
where $a$ $(a^{\dagger})$ is the annihilation (creation) operator of the cavity field with the resonance frequency $\omega_{c}$, and the $g$ term describes the interaction between the cavity mode and the vibrational mode.

It can be seen from Hamiltonian~(\ref{Hamilthh}) that the conditional displacement of the vibrational mode in the phase space is determined by the ratio of the conditional coupling strength $\lambda$ over the resonance frequency $\omega_{\nu}$ of the vibrational mode. In current experiments, the value of the ratio $\lambda/\omega_{\nu}$ is much smaller than $1$, which means that the displacement of the vibrational mode will be shadowed by its zero-point fluctuation, and that the distance between these superposition coherent-state components are too small to be distinguished from each other. Consequently, how to generate macroscopic cat states with distinguishable superposition components becomes an important task. Motivated by a scheme for generation of macroscopic cat states superposed by two distinct coherent states~\cite{Liao2016}, in this paper, we propose the Floquet-sideband-assisted displacement scheme to generate macroscopic entangled cat states by introducing proper frequency modulations to the cavity field and the vibrational mode. The Hamiltonian describing the frequency modulations reads
\begin{equation}
H_{m}(t)=\xi\omega_{0}\cos(\omega_{0}t)(a^{\dag}a+b^{\dag}b),\label{modulated}
\end{equation}
where $\omega_{0}$ is the modulation frequency and $\xi$ is the dimensionless modulation amplitude. The action of the frequency modulation is to enhance the vibrational displacement with the Floquet-sideband modes. By choosing proper modulation parameters, one of these sideband modes can compensate the vibrational driving detuning and enhance the vibrational displacement.

Based on the above analyses, the total Hamiltonian of the system can be written as
\begin{equation}
H(t) =H_{e-v}+H_{c-v}+H_{m}(t),\label{Hamilt1}
\end{equation}
where $H_{e-v}$, $H_{c-v}$, and $H_{m}(t)$ have been defined in Eqs.~(\ref{Hamilthh}), (\ref{Hamilthc}), and (\ref{modulated}), respectively. Below, we will elaborate how to enhance the vibrational displacement by analyzing the action of the Floquet-sideband modes induced by the frequency modulations.

In order to study the influence of the frequency modulations of the cavity mode and the vibrational mode on the dynamics of the system, we perform a transformation defined by the operator
\begin{equation}
T_{1}(t) =\exp[-i\xi\sin(\omega_{0}t)(a^{\dag}a+b^{\dag}b)] \label{transfV}
\end{equation}
on the Hamiltonian $H(t)$ in Eq.~(\ref{Hamilt1}). The transformed Hamiltonian becomes
\begin{eqnarray}
H_{1}(t)&=&T^{\dag}_{1}(t)H(t)T_{1}(t)-iT^{\dag}_{1}(t)\dot{T}_{1}(t) \nonumber\\
&=&\omega_{e}\sigma_{+}\sigma_{-}+\omega_{v}b^{\dag}b+\omega_{c}a^{\dag}a+g(a^{\dag}b+ab^{\dag}) \nonumber\\
&&+\lambda\sigma_{+}\sigma_{-}(b^{\dag}e^{i\xi\sin(\omega_{0}t)}+be^{-i\xi\sin(\omega_{0}t)}). \label{Hamilt2}
\end{eqnarray}
Here, the exponentials of sinusoidal function will introduce the Floquet-sideband modes, which can enhance the conditional displacement of the molecular vibration mode.

To diagonalize the $g$ term in Eq.~(\ref{Hamilt2}), we introduce the transformation operator $T_{2}=\exp[\vartheta(a^{\dag}b-ab^{\dag})]$, then we obtain the transformed Hamiltonian
\begin{eqnarray}
H_{2}(t)&=&T^{\dag}_{2}H_{1}(t)T_{2} \nonumber\\
&=&\omega_{e}\sigma_{+}\sigma_{-}+\omega_{+}b^{\dag}b+\omega_{-}a^{\dag}a+\lambda\sigma_{+}\sigma_{-}  \nonumber\\
&&\times[(b^{\dag}\cos\vartheta-a^{\dag}\sin\vartheta)e^{i\xi\sin(\omega_{0}t)}+\text{H.c.}],\label{defH2}
\end{eqnarray}
where we introduce the resonance frequencies of the two hybrid modes as
\begin{equation}
\omega_{\pm}=\frac{\omega_{c}+\omega_{v}}{2}\pm\frac{\omega_{v}-\omega_{c}}{2}\cos(2\vartheta)\pm g\sin(2\vartheta).
\end{equation}
The mixing angle in Eq.~(\ref{defH2}) is defined by
\begin{equation}
\vartheta=\frac{1}{2}\arctan\left(\frac{2g}{\omega_{m}-\omega_{c}}\right).
\end{equation}

In a rotating frame defined by the transformation operator $T_{3}(t)=\exp(-iH_{0}t)$ with $H_{0}=\omega_{e}\sigma_{+}\sigma_{-}+\omega_{+}b^{\dag}b+\omega_{-}a^{\dag}a$, the transformed Hamiltonian becomes
\begin{equation}
H_{3}(t)=T^{\dag}_{3}(t)[H_{2}(t)-H_{0}]T_{3}(t)=H_{a}(t)+H_{b}(t),\label{Hamiltilde}
\end{equation}
where
\begin{subequations}
\label{Hamiltab}
\begin{align}
H_{a}(t)=&-\lambda\sin(\vartheta)\sigma_{+}\sigma_{-}(a^{\dag}e^{i[\omega_{-}t+\xi\sin(\omega_{0}t)]}+\text{H.c.}),\label{Hamilta}\\
H_{b}(t)=&\lambda\cos(\vartheta)\sigma_{+}\sigma_{-}(b^{\dag}e^{i[\omega_{+}t+\xi\sin(\omega_{0}t)]}+\text{H.c.}),\label{Hamiltb}
\end{align}
\end{subequations}
are the interaction Hamiltonians associated with the cavity mode and the vibrational mode, respectively. Note that Hamiltonian~(\ref{Hamiltilde}) can be divided into two parts because of $[H_{a}(t),H_{b}(t)]=0$. Below, we will treat the two Hamiltonians $H_{a}(t)$ and $H_{b}(t)$ separately.

The action of the Floquet-sideband modes can be seen by expanding the functions $\exp\{\pm i[\omega_{\pm}t+\xi\sin(\omega_{0}t)]\}$ in Eq.~(\ref{Hamiltab}) with the Jacobi-Anger expansion
\begin{equation}
\exp[i\xi\sin(\omega_{0}t)]=\sum_{n=-\infty}^{\infty}J_{n}(\xi)e^{in\omega_{0}t},
\end{equation}
then the Hamiltonian $H_{a}(t)$ can be written as
\begin{eqnarray}
H_{a}(t)&=&-\lambda\sin(\vartheta)\sigma_{+}\sigma_{-}\left[J_{-1}(\xi)(a^{\dag}e^{i(\omega_{-}-\omega_{0})t}+\text{H.c.})+\cdots\right. \nonumber\\
&&\left.\underline{+J_{-n_{a}}(\xi)(a^{\dag}e^{i(\omega_{-}-n_{a}\omega_{0})t}+\text{H.c.})}+\cdots \right.\nonumber\\
&&\left.+J_{-n}(\xi)(a^{\dag}e^{i(\omega_{-}-n\omega_{0})t}+\text{H.c.})+\cdots \right.\nonumber\\
&&\left.+\sum_{n=0}^{\infty}J_{n}(\xi)(a^{\dag}e^{i(\omega_{-}+n\omega_{0})t}+\text{H.c.})\right].\label{HsummationHa}
\end{eqnarray}
where $J_{n}(x)$ is the Bessel function of the first kind, with $n$ being an integer. It can be seen from Eq.~(\ref{HsummationHa}) that those terms with different frequencies $n\omega_{0}$ correspond to different Floquet-sideband modes. For the motivation of generation of macroscopical entangled cat states, we choose the proper parameters such that the conditions
\begin{equation}
\{\omega_{-}, \omega_{0}\}\gg\lambda>\lambda|\sin(\vartheta)J_{n}(\xi)| \label{paracondaa}
\end{equation}
are satisfied. Furthermore, we choose another proper parameters such that there is one near-resonant term in Eq.~(\ref{HsummationHa}). Without loss of generality, we denote $n_{a}$ as the characteristic number corresponding to the near-resonant term (i.e., the underlined term, the target term), then the effective driving detuning of the target term is
\begin{equation}
\delta_{a}=\omega_{-}-n_{a}\omega_{0}.\label{deltaa}
\end{equation}

For a given characteristic number $n_{a}$, we choose a appropriate modulation frequency $\omega_{0}$ such that $\delta_{a}$ can be comparable or even smaller than the coupling coefficient $\lambda$, whereas all other terms are high-frequency-oscillating terms that satisfy the parameter condition (\ref{paracondaa}). The high-frequency-oscillating terms can be neglected by the rotating-wave approximation (RWA), but the near-resonant term (corresponding to a characteristic number $n_{a}$) with the form of $-\lambda\sin(\vartheta)J_{-n_{a}}(\xi)\sigma_{+}\sigma_{-}(a^{\dag}e^{i\delta_{a}t}+ae^{-i\delta_{a}t})$ should be kept. Then the Hamiltonian $H_{a}(t)$ can be approximated as
\begin{equation}
H_{a}(t)\approx-g_{a}\sigma_{+}\sigma_{-}(a^{\dag}e^{i\delta_{a}t}+ae^{-i\delta_{a}t}), \label{Hamappra}
\end{equation}
where $g_{a}=\lambda\sin(\vartheta)J_{-n_{a}}(\xi)$ is the effective coupling strength.

Similarly, the Hamiltonian $H_{b}(t)$ can be written with the Jacobi-Anger expansion as
\begin{eqnarray}
H_{b}(t)&=&\lambda\cos(\vartheta)\sigma_{+}\sigma_{-}\left[J_{-1}(\xi)(b^{\dag}e^{i(\omega_{+}-\omega_{0})t}+\text{H.c.})+\cdots\right. \nonumber\\
&&\left.\underline{+J_{-n_{b}}(\xi)(b^{\dag}e^{i(\omega_{+}-n_{b}\omega_{0})t}+\text{H.c.})}+\cdots \right.\nonumber\\
&&\left.+J_{-n}(\xi)(b^{\dag}e^{i(\omega_{+}-n\omega_{0})t}+\text{H.c.})+\cdots \right.\nonumber\\
&&\left.+\sum_{n=0}^{\infty}J_{n}(\xi)(b^{\dag}e^{i(\omega_{+}+n\omega_{0})t}+\text{H.c.})\right].\label{HsummationHb}
\end{eqnarray}
By choosing proper parameters such that the conditions
\begin{equation}
\{\omega_{+}, \omega_{0}\}\gg\lambda>\lambda|\cos(\vartheta)J_{n}(\xi)| \label{paracondab}
\end{equation}
are satisfied and the sideband with the index $n_{b}$ to be near resonant, then we can obtain the following approximate Hamiltonian by making the RWA as
\begin{equation}
H_{b}(t)\approx g_{b}\sigma_{+}\sigma_{-}(b^{\dag}e^{i\delta_{b}t}+be^{-i\delta_{b}t}), \label{Hamapprb}
\end{equation}
where $g_{b}=\lambda\cos(\vartheta)J_{-n_{b}}(\xi)$ is the effective coupling strength and $\delta_{b}=\omega_{+}-n_{b}\omega_{0}$ is the detuning.

\begin{figure}[tbp]
\center
\includegraphics[ width=0.47 \textwidth]{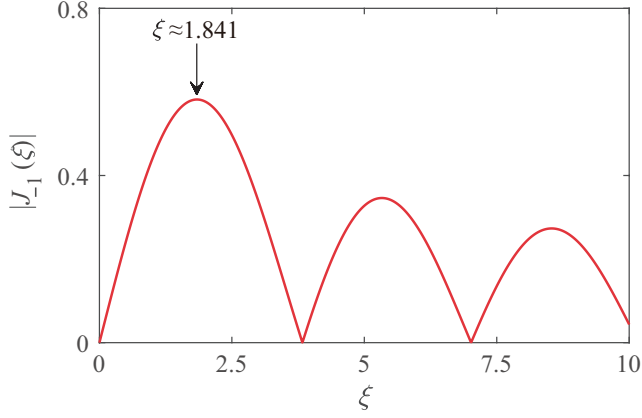}
\caption{The absolute value of the Bessel function $|J_{-1}(\xi)|$ of the first kind as a function of $\xi$. The first peak value of $|J_{-1}(\xi)|$ is obtained at $\xi\approx1.841$.}
\label{Fig2}
\end{figure}

The Hamiltonians $H_{a}(t)$ and $H_{b}(t)$ describe that the two bosonic modes (the cavity mode and the vibrational mode) are conditionally displaced by the electronic degree of freedom. Only when the electronic degree of freedom is prepared in the excited state $|e\rangle$, the two bosonic modes are displaced. If the molecule is initially prepared in a coherent superposition of the ground state $|g\rangle$ and the excited state $|e\rangle$, the system will be evolved into entangled states. Moreover, it can be seen from Eqs.~(\ref{Hamappra}) and~(\ref{Hamapprb}) that the magnitude of $|g_{a}|$ and $|g_{b}|$ can be maximized by optimizing the value of $\xi$. For below convenience, we choose $n_{a}=1$, $n_{b}=1$, and $\xi=1.841$ for reaching large values of $|J_{-n_{a}}(\xi)|$ and $|J_{-n_{b}}(\xi)|$ simultaneously (see Fig.~\ref{Fig2}). We also choose small values of $\delta_{a}$ and $\delta_{b}$ by adjusting $\omega_{0}$ to strongly enhance the displacement of the two bosonic modes. The conditional dynamic can be used to create macroscopic superpositions of large-amplitude coherent states. The physical mechanism of the entangled cat-state generation can be seen more clearly by writing the approximate Hamiltonian of the system as the following conditional coupling form
\begin{equation}
H_{\mbox{\scriptsize RWA}}(t)=H_{e}(t)\otimes\vert e\rangle\langle e\vert +H_{g}\otimes\vert g\rangle\langle g\vert, \label{Hamiltrwa}
\end{equation}
where $H_{e}(t)=g_{b}(b^{\dag}e^{i\delta_{b}t}+\text{H.c.})-g_{a}(a^{\dag}e^{i\delta_{a}t}+\text{H.c.})$ and $H_{g}=0$ are the effective Hamiltonians corresponding to the states $|e\rangle$ and $|g\rangle$, which govern the evolution of the cavity field and the vibrational mode.

\section{Generation of entangled cat states\label{sec:stategeneration}}

In this section, we study how to generate macroscopic entangled cat states based on the dynamical evolution of the coupled system. We also study the quantum interference and coherence effects of the generated entangled cat states by analyzing the joint Wigner function and the degree of entanglement.

\subsection{Analytical solution under the RWA\label{ssec:analytical}}

To clearly see the results of the state generation, we first calculate the analytical expression of the generated states with the approximate Hamiltonian~(\ref{Hamiltrwa}). In the derivation of the approximate Hamiltonian~(\ref{Hamiltrwa}), we have made three steps of transformations. To know these state transformations, we summarize the relations between these transformed Hamiltonians as
\begin{subequations}
\label{Hamiltilsteps}
\begin{align}
H_{1}(t)=&T^{\dag}_{1}(t)H(t)T_{1}(t)-iT^{\dag}_{1}(t)\dot{T}_{1}(t),\\
H_{2}(t)=&T^{\dag}_{2}H_{1}(t)T_{2},\\
H_{3}(t)=&T^{\dag}_{3}(t)[H_{2}(t)-H_{0}]T_{3}(t).
\end{align}
\end{subequations}
Accordingly, the relations between the states in these three transformed representations are given by
\begin{subequations}
\label{transformation}
\begin{align}
\vert\psi(t)\rangle=&T_{1}(t)\vert\varphi(t)\rangle,\\
\vert\varphi(t)\rangle=&T_{2}\vert\Phi(t)\rangle,\\
\vert \Phi(t)\rangle=&T_{3}(t)\vert\phi(t)\rangle.
\end{align}
\end{subequations}
In the following, we will calculate the evolution of the system in the representation associated with the state $\vert\phi(t)\rangle$. The state $\vert\psi(t)\rangle$ in the Schr\"{o}dinger representation can then be obtained via the transformations given in Eq.~(\ref{transformation}). In particular, the evolution of the state $\vert\phi(t)\rangle$ is governed by the Schr\"{o}dinger equation
\begin{equation}
i\frac{\partial}{\partial t}\vert\phi(t)\rangle=H_{3}(t)\vert\phi(t)\rangle\approx H_{\mbox{\scriptsize RWA}}(t)\vert\phi(t)\rangle. \label{statesteps}
\end{equation}

At the initial moment, we have the relations $T_{1}(0)=T_{3}(0)=I$ and $\vert\psi(0)\rangle=T_{1}(0)\vert\varphi(0)\rangle =T_{1}(0)T_{2}\vert \Phi(0)\rangle =T_{1}(0)T_{2}T_{3}(0)\vert\phi(0)\rangle$, then $\vert\phi(0)\rangle=T^{\dag}_{2}\vert\psi(0)\rangle$ can be obtained. We consider the initial state of the system in the original representation as
\begin{equation}
\vert\psi(0)\rangle=\frac{1}{\sqrt{2}}(\vert g\rangle+\vert e\rangle)\vert 0\rangle_{a}\vert 0\rangle_{b},
\end{equation}
which leads to $\vert\phi(0)\rangle=(\vert g\rangle+\vert e\rangle)\vert 0\rangle_{a}\vert 0\rangle_{b}/\sqrt{2}$. According to the time evolution operator
\begin{equation}
U_{\mbox{\scriptsize RWA}}(t)=\mathcal{T}\exp\left[-i\int_{0}^{t}H_{\mbox{\scriptsize RWA}}(\tau)d\tau\right],\label{unitaryRWA}
\end{equation}
with ``$\mathcal{T}$" being the time-ordering operator, the state of the system at time $t$ becomes
\begin{eqnarray}
\vert\phi(t)\rangle &=&U_{\textrm{RWA}}(t)\vert\phi(0)\rangle  \nonumber\\
&=&\frac{1}{\sqrt{2}}(U_{e}(t)\vert e\rangle\vert 0\rangle_{a}\vert 0\rangle_{b}+U_{g}\vert g\rangle\vert 0\rangle_{a}\vert 0\rangle_{b}),\label{unitevolution}
\end{eqnarray}
where $U_{e}(t)=\mathcal{T}\exp[-i\int_{0}^{t}H_{e}(\tau)d\tau]$ and $U_{g}=I$. Since the value of the commutator $[H_{e}(t_{1}),H_{e}(t_{2})]$ is a ``c"-number, the unitary evolution operator associated with the Hamiltonian $H_{e}(t)$ can be obtained by using the Magnus expansion as (see the Appendix)
\begin{eqnarray}
U_{e}(t)&=&U_{a}(t)U_{b}(t) \nonumber\\
&=&\exp[i\theta_{a}(t)]\exp[\eta(t)a-\eta^{\ast}(t)a^{\dag}] \nonumber\\
&&\times\exp[i\theta_{b}(t)]\exp[\zeta(t)b^{\dag}-\zeta^{\ast}(t)b],\label{unitary}
\end{eqnarray}
where we introduce the phase factors
\begin{subequations}
\begin{align}
\theta_{a}(t)=&\left(\frac{g_{a}}{\delta_{a}}\right)^{2}[\delta_{a}t-\sin(\delta_{a}t)],\\
\theta_{b}(t)=&\left(\frac{g_{b}}{\delta_{b}}\right)^{2}[\delta_{b}t-\sin(\delta_{b}t)],
\end{align}
\end{subequations}
and the displacement amplitudes
\begin{subequations}
\begin{align}
\eta(t)=&\frac{g_{a}}{\delta_{a}}(1-e^{i\delta_{a}t}), \\
\zeta(t)=&\frac{g_{b}}{\delta_{b}}(1-e^{i\delta_{b}t}),
\end{align}
\end{subequations}
for the cavity mode and the vibrational mode, respectively. By utilizing the unitary evolution operator~(\ref{unitary}), the state of the system at time $t$ can be obtained as
\begin{equation}
\vert\phi(t)\rangle =\frac{1}{\sqrt{2}}\left(e^{i[\theta_{a}(t)+\theta_{b}(t)]}\vert e\rangle\vert-\eta(t)\rangle_{a}\vert\zeta(t)\rangle_{b}+\vert g\rangle\vert0\rangle_{a}\vert 0\rangle_{b}\right).
\end{equation}

By performing the transformation $\vert\psi(t)\rangle =T_{1}(t)T_{2}T_{3}(t)\vert\phi(t)\rangle$, the corresponding state in the original representation can be expressed as
\begin{equation}
\vert\psi(t)\rangle=\frac{1}{\sqrt{2}}\left(e^{i\theta(t)}\vert e\rangle\vert\alpha(t)\rangle_{a}\vert\beta(t)\rangle _{b}+\vert g\rangle\vert 0\rangle_{a}\vert 0\rangle_{b}\right), \label{originalstatepsi}
\end{equation}
where $\theta(t)=\theta_{a}(t)+\theta_{b}(t)-\omega_{e}t$ is a phase factor and the coherent-state amplitudes are given by
\begin{subequations}
\label{alphabeta}
\begin{align}
\alpha(t)=&\frac{g_{b}\sin(\vartheta)}{\delta_{b}}(1-e^{i\delta_{b}t})e^{-i[\xi\sin(\omega_{0}t)+\omega_{+}t]} \nonumber\\
&-\frac{g_{a}\cos(\vartheta)}{\delta_{a}}(1-e^{i\delta _{a}t})e^{-i[\xi\sin(\omega_{0}t)+\omega_{-}t]}\label{amplitudea},\\
\beta(t)=&\frac{g_{b}\cos(\vartheta)}{\delta_{b}}(1-e^{i\delta_{b}t})e^{-i[\xi\sin(\omega_{0}t)+\omega_{+}t]}\nonumber\\
&+\frac{g_{a}\sin(\vartheta)}{\delta_{a}}(1-e^{i\delta _{a}t})e^{-i[\xi\sin(\omega_{0}t)+\omega_{-}t]}.\label{amplitudeb}
\end{align}
\end{subequations}
For the state $\vert\psi(t)\rangle$, the average excitation numbers in the cavity mode and the vibrational mode are given by
\begin{subequations}
\label{aveexcitANAB}
\begin{align}
\langle\psi(t)\vert a^{\dag}a\vert\psi(t)\rangle=&\frac{1}{2}|\alpha(t)|^2, \label{aveexcitANA}\\
\langle\psi(t)\vert b^{\dag}b\vert\psi(t)\rangle=&\frac{1}{2}|\beta(t)|^2.\label{aveexcitANB}
\end{align}
\end{subequations}

By expressing the states of the electronic degree of freedom with the bases $\vert\pm\rangle=(\vert e\rangle\pm\vert g\rangle)/\sqrt{2}$, the generated state $\vert\psi(t)\rangle$ can be expressed as
\begin{equation}
\vert\psi(t)\rangle=\frac{1}{2\mathcal{M}_{+}(t)}\vert+\rangle\vert\psi_{+}(t)\rangle+\frac{1}{2\mathcal{M}_{-}(t)}\vert-\rangle\vert\psi_{-}(t)\rangle,\label{totalpsistate}
\end{equation}
where we introduced the entangled cat states for the two modes as~\cite{Vitali2000,Chai1992}
\begin{equation}
\vert\psi_{\pm}(t)\rangle=\mathcal{M}_{\pm}(t)(e^{i\theta(t)}\vert\alpha(t)\rangle_{a}\vert\beta(t)\rangle_{b}\pm\vert 0\rangle_{a}\vert 0\rangle_{b}),\label{psistate}
\end{equation}
with the normalization constants
\begin{equation}
\mathcal{M}_{\pm}(t)=\frac{1}{\sqrt{2}}\left(1\pm\cos[\theta(t)]e^{-\frac{1}{2}(\vert\alpha(t)\vert^{2}+\vert\beta(t)\vert^{2})}\right)^{-1/2}.
\end{equation}

It follows from Eq.~(\ref{totalpsistate}) that, by performing a measurement of the electronic states in the bases $|\pm\rangle$, the two bosonic modes $a$ and $b$ will collapse into the entangled cat states $|\psi_{\pm}(t)\rangle$. The corresponding probabilities for detection of the electronic states in $|\pm\rangle$ are
\begin{equation}
\mathcal{P}_{\pm}(t)=\frac{1}{4\mathcal{M}^{2}_{\pm}(t)}=\frac{1}{2}\left(1\pm\cos[\theta(t)]e^{-\frac{1}{2}(\vert\alpha(t)\vert^{2}+\vert\beta(t)\vert^{2})}\right),\label{proapproximate}
\end{equation}
which represent the success probabilities for the generation of macroscopic entangled cat states in the ideal case.

\subsection{Entanglement between the cavity field and the vibrational mode}

The degree of entanglement of the generated entangled cat states can be quantized by calculating the logarithmic negativity~\cite{Vidal2002,Plenio2005}. For a two-partite system described by the density matrix $\rho$, the logarithmic negativity is defined by
\begin{equation}
N=\log_{2}\left\Vert\rho^{T_{b}}\right\Vert_{1},\label{lognegtivity}
\end{equation}
where ``$T_{b}$" denotes the partial transpose of the density matrix $\rho$ with respect to the vibrational mode $b$, and the trace norm $\left\Vert\rho^{T_{b}}\right\Vert_{1}$ is defined by
\begin{equation}
\left\Vert\rho^{T_{b}}\right\Vert_{1}=\textrm{Tr}\left[\sqrt{(\rho^{T_{b}})^{\dag}\rho^{T_{b}}} \right].\label{tracenorm}
\end{equation}

\begin{figure}[tbp]
\center
\includegraphics[ width=0.47 \textwidth]{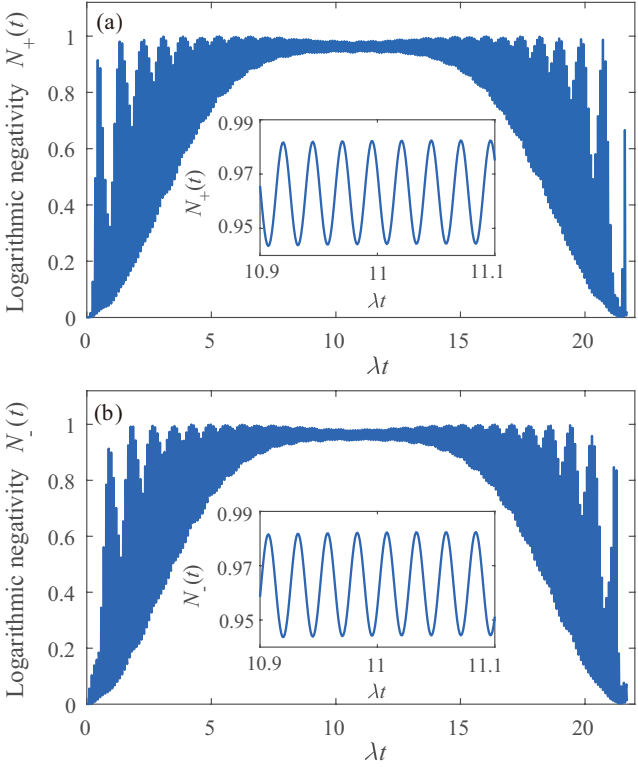}
\caption{Dynamics of the logarithmic negativities (a) $N_{+}(t)$ and (b) $N_{-}(t)$ for the states $\vert\psi_{\pm}(t)\rangle$ in Eq.~(\ref{psistate}), respectively. The insets are zoomed-in plots of the logarithmic negativities in the middle duration of one period. Here, we set the detuning to be $\delta_{a}=-g_{a}/2$, and other parameters are given by $\omega_{c}/\lambda=50$, $\omega_{m}/\lambda=50.5$, $\omega_{e}/\lambda=250$, $g/\lambda=2.5$, and $\xi=1.841$. }
\label{Fig3}
\end{figure}
Below, we will calculate the logarithmic negativity of the generated states $\vert\psi_{\pm}(t)\rangle$, which can be expanded in the Fock space as
\begin{equation}
\vert\psi_{\pm}(t)\rangle=\mathcal{M}_{\pm}(t)\sum_{m,k=0}^{\infty}C^{\pm}_{m,k}(t)\vert m\rangle_{a}\vert k\rangle_{b},\label{fockstate}
\end{equation}
where we introduce the probability amplitudes
\begin{equation}
C^{\pm}_{m,k}(t)=\frac{e^{i\theta(t)-\frac{1}{2}(\vert\beta(t)\vert^{2}+\vert\alpha(t)\vert^{2})}}{\sqrt{m!k!}}\alpha^{m}(t)\beta^{k}(t)\pm\delta_{m,0}\delta_{0,k}.\label{amplitude}
\end{equation}
Using Eqs.~(\ref{lognegtivity}) and~(\ref{tracenorm}), we can calculate numerically the logarithmic negativity of entangled cat states $\vert\psi_{\pm}(t)\rangle$.

In Fig.~\ref{Fig3}, we show the dynamics of the logarithmic negativities $N_{\pm}(t)$ for the states $\vert\psi_{\pm}(t)\rangle$. We find that the logarithmic negativities $N_{\pm}(t)$ exhibit fast oscillations in a whole period, which is caused by the phase factor $\theta(t)$. At time $t=2\pi/\delta_{a}$ (the end of one period), the coherent amplitudes $|\alpha(t)|=0$ and $|\beta(t)|\approx0$, which means that the molecular vibrational mode and the cavity-field mode decouple with each other and the entanglement disappears. In the middle duration of a period, the amplitude of the oscillation decreases, and the logarithmic negativities $N_{\pm}(t)$ become approximately stable and reach the maximum values (see the insert diagram). This is because the coherent amplitude $|\alpha(t)|$ and $|\beta(t)|$ are large enough such that $\vert \langle0\vert\alpha(t)\rangle\vert^2=\exp[-|\alpha(t)|^{2}]\approx0$ and $\vert \langle0\vert\beta(t)\rangle\vert^2=\exp[-|\beta(t)|^{2}]\approx0$, then the logarithmic negativities can reach maximum values around $1$.

\subsection{The joint Wigner function of the entangled cat states}

\begin{figure}[tbp]
\center
\includegraphics[ width=0.47 \textwidth]{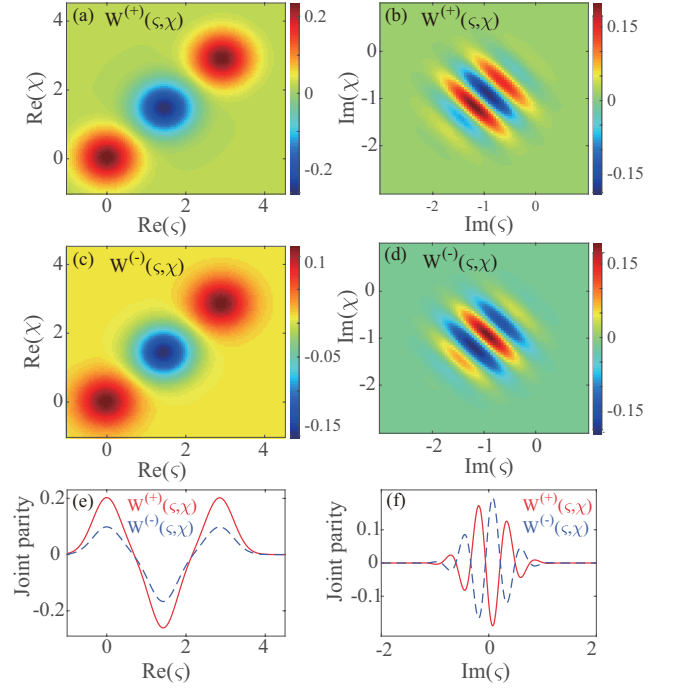}
\caption{Plane cuts of the two-mode joint Wigner functions $W^{(\pm)}(\varsigma,\chi)$ along (a),(c) the Re$(\varsigma)$-Re$(\chi)$ axes and (b),(d) the Im$(\varsigma)$-Im$(\chi)$ axes for the entangled cat states with $\vert\alpha(t_{s})\vert\approx2.88$ and $\vert\beta(t_{s})\vert\approx2.82$. (e) The diagonal line-cuts of (a) and (c) corresponding to the plots of the joint Wigner functions $W^{(+)}(\varsigma,\chi)$ (red solid curve) and $W^{(-)}(\varsigma,\chi)$ (blue dashed curve) along $\text{Re}(\varsigma)=\text{Re}(\chi)$ with $\text{Im}(\varsigma)=0$ and $\text{Im}(\chi)=0.6$. (f) The diagonal line-cuts of (b) and (d), corresponding to the plots of the joint Wigner functions $W^{(+)}(\varsigma,\chi)$ (red solid curve) and $W^{(-)}(\varsigma,\chi)$ (blue dashed curve) along $\text{Im}(\varsigma)=\text{Im}(\chi)$ with $\text{Re}(\varsigma)=\text{Re}(\chi)=1$.}
\label{Fig4}
\end{figure}

To exhibit quantum interference and coherence effects of the generated entangled cat states, we calculate the joint Wigner function of the cavity-field mode and the molecular vibrational mode~\cite{Wang2016,Zhong2018}. A full quantum state tomography of the two-mode system can be realized by measuring the joint Wigner function. For a two-bosonic-mode ($a$ and $b$) system, the joint Wigner function is defined by
\begin{equation}
W(\varsigma,\chi)=\frac{4}{\pi^{2}}\left\langle D_{a}(\varsigma)(-1)^{a^{\dag}a}D_{a}^{\dag}(\varsigma)D_{b}(\chi)(-1)^{b^{\dag}b}D_{b}^{\dag }(\chi) \right\rangle,\label{jointwigf}
\end{equation}
where $D_{a}(\varsigma)=\exp(\varsigma a^{\dag}-\varsigma^{\ast}a)$ and $D_{b}(\chi)=\exp (\chi b^{\dag}-\chi^{\ast}b)$ are, respectively, the displacement operators of the cavity mode and the vibrational mode, with $\varsigma$ and $\chi$ being the complex parameters defining the coordinates in the joint phase space. Equation (\ref{jointwigf}) shows that $W(\varsigma,\chi)$ is a real function in the $4$D phase space [Re$(\varsigma)$, Im$(\varsigma)$, Re$(\chi)$, and Im$(\chi)$]. The value of $W(\varsigma,\chi)$ (after rescaling by $4/\pi^{2}$) at each point can be measured from the expectation value of the joint parity operator $(-1)^{a^{\dag}a+b^{\dag}b}$ transformed by the displacements $D_{a}(\varsigma)D_{b}(\chi)$. The joint Wigner function can be used to demonstrate nonclassical property between the cavity field and the molecular vibration. In order to explain the core features in these $4$D Wigner functions of the states $\vert\psi_{\pm}(t)\rangle$, we show the $2$D cuts of the Wigner functions along both the Re$(\varsigma)$-Re$(\chi)$ plane and the Im$(\varsigma)$-Im$(\chi)$ plane.

In Figs.~\ref{Fig4}(a) and~\ref{Fig4}(c), we plot the plane cut of the two-mode Wigner functions $W_{\pm}(\varsigma,\chi)$ along the Re$(\varsigma)$-Re$(\chi)$ axes. Each of the figures contains two positive-valued Gaussian balls, which represent the probability distribution of the two superposed two-mode coherent states. Meanwhile, there is a negative-valued Gaussian ball around the center between the two balls, which is a signature of quantum interference effect between the two superposition components. In Figs.~\ref{Fig4}(b) and~\ref{Fig4}(d), we can see the interference pattern of the coherent superposition along the Im$(\varsigma)$-Im$(\chi)$ axes, which is a result of quantum interference between two quasi-classical states $\vert\alpha(t)\rangle_{a}\vert\beta(t)\rangle_{b}$ and $\vert 0\rangle_{a}\vert 0\rangle_{b}$. Moreover, the interference pattern is similar to that for the single-mode cat state, and the interference features associated with the two entangled cat states $\vert\psi_{\pm}(t)\rangle$ are complementary. This feature can be seen more clearly in the diagonal line-cuts of (a-d), as shown in Figs.~\ref{Fig4}(e) and~\ref{Fig4}(f).

\subsection{The generated states under the full Hamiltonian\label{ssec:exact state}}

The exact state evolution of this system can be calculated by numerically solving the Schr\"{o}dinger equation under the full Hamiltonian~(\ref{Hamilt1}). To this end, we express a general pure state of the system as
\begin{equation}
\vert\Psi(t)\rangle =\sum_{n,j=0}^{\infty}(A_{n,j}(t)\vert e\rangle\vert n\rangle_{a}\vert j\rangle_{b}+B_{n,j}(t)\vert g\rangle\vert n\rangle_{a}\vert j\rangle_{b}),\label{Psinumexa}
\end{equation}
where $\vert n\rangle_{a}$ and $\vert j\rangle_{b}$ are the Fock states of the cavity mode and the molecular vibrational mode, respectively. By substituting the state $\vert\Psi(t)\rangle$ in Eq.~(\ref{Psinumexa}) and the full Hamiltonian (\ref{Hamilt1}) into the Schr\"{o}dinger equation, we obtain the equations of motion for these probability amplitudes $A_{n,j}(t)$ and $B_{n,j}(t)$ as
\begin{subequations}
\begin{align}
\dot{A}_{n,j}(t)=&-iA_{n,j}(t)[\omega_{e}+n\omega_{c}+j\omega_{m}+(n+j)\xi\omega_{0}\cos(\omega_{0}t)]\nonumber\\
&-ig_{0}[\sqrt{n(j+1)}A_{n-1,j+1}(t)+\sqrt{(n+1)j}A_{n+1,j-1}(t)]\nonumber\\
&-ig_{1}[\sqrt{j}A_{n,j-1}(t)+\sqrt{(j+1)}A_{n,j+1}(t)],\label{eqofmotproampa}\\
\dot{B}_{n,j}(t)=&-iB_{n,j}(t)[n\omega_{c}+j\omega_{m}+(n+j)\xi\omega_{0}\cos(\omega_{0}t)]\nonumber\\
&-ig_{0}[\sqrt{n(j+1)}B_{n-1,j+1}(t)+\sqrt{(n+1)j}B_{n+1,j-1}(t)],\label{eqofmotproampb}
\end{align}
\end{subequations}
where $n$ and $j$ are natural numbers. Corresponding to the initial state $\vert+\rangle\vert 0\rangle_{a}\vert 0\rangle_{b}$, the initial conditions of these probability amplitudes are $A_{n,j}( 0)=B_{n,j}(0)=\delta_{n,0}\delta_{j,0}/\sqrt{2}$. Combining with these initial conditions, the evolution of these probability amplitudes $A_{n,j}$ and $B_{n,j}$ can be obtained by numerically solving Eqs.~(\ref{eqofmotproampa}) and (\ref{eqofmotproampb}).
\begin{figure}[tbp]
\center
\includegraphics[ width=0.47 \textwidth]{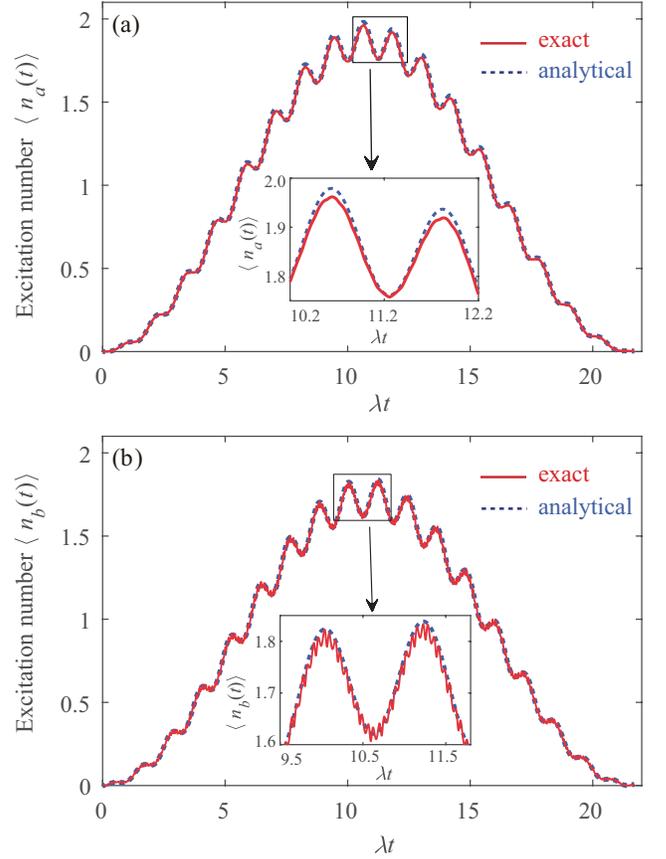}
\caption{Time dependence of the average excitation numbers (a) $\langle n_{a}(t)\rangle$ of the cavity mode and (b) $\langle n_{b}(t)\rangle$ of the vibrational mode. The two curves in each figure are based on the exact result (red) at selected frequency $\omega_{c}/\lambda=50$ and the analytical result (blue) given in Eq.~(\ref{aveexcitANAB}) , respectively. The insets are zoomed-in plots of the average excitation numbers in the middle duration of one period.
Other parameters used are the same as those in Fig~\ref{Fig3}.}
\label{Fig5}
\end{figure}

Accordingly, we can calculate the average excitation numbers $\langle n_{a}(t)\rangle$ and $\langle n_{b}(t)\rangle$ of the cavity mode and the vibrational mode, respectively. For the state $|\Psi(t)\rangle$, the average excitation numbers can be obtained as
\begin{subequations}
\begin{align}
\langle n_{a}(t)\rangle=&\sum_{n,j=0}^{\infty}[n(|A_{n,j}(t)|^{2}+|B_{n,j}(t)|^{2})],\label{avnaclonumer}\\
\langle n_{b}(t)\rangle=&\sum_{n,j=0}^{\infty}[j(\vert A_{n,j}(t)\vert^{2}+\vert B_{n,j}(t)\vert^{2})].\label{avnbclonumer}
\end{align}
\end{subequations}

In Figs.~\ref{Fig5}(a) and~\ref{Fig5}(b), we display the time dependence of the average excitation numbers $\langle n_{a}(t)\rangle$ and $\langle n_{b}(t)\rangle$ at selected value of $\omega_{c}/\lambda=50$, which are compared with the analytical results given by Eq.~(\ref{aveexcitANAB}). From these curves we can see that the peak values of the average excitation numbers are located around time $\pi/\delta_{a}$ and the numerical results matches well with the analytical results. This result can be well explained by Eqs.~(\ref{alphabeta}) and~(\ref{aveexcitANAB}), the displacements of the cavity mode and the vibrational mode reach their maximum values at the detection time.

\begin{figure}[tbp]
\center
\includegraphics[ width=0.47 \textwidth]{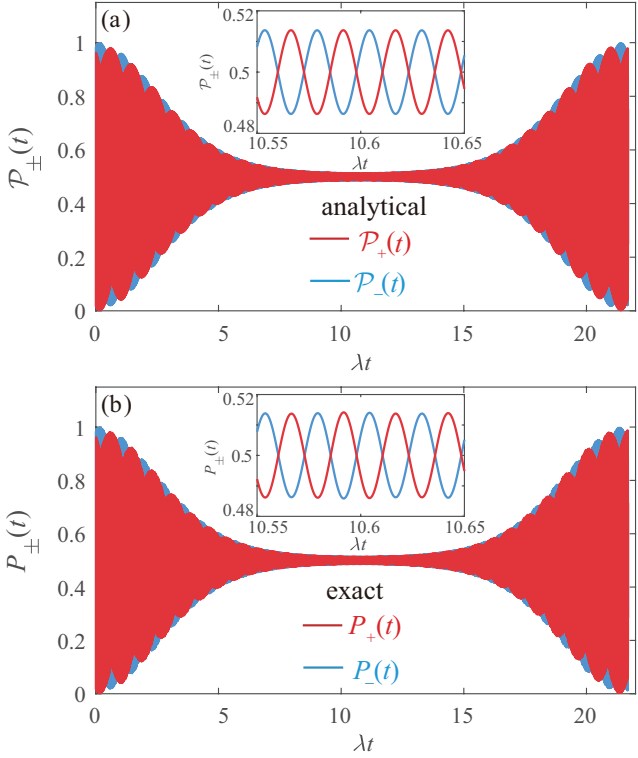}
\caption{Plots of (a) the analytical detection probabilities $\mathcal{P}_{\pm}(t)$ given in Eq.~(\ref{proapproximate}) and (b) the exact detection probabilities $P_{\pm}(t)$ given in Eq.~(\ref{proclose}) [with the selected vaule $\omega_{c}/\lambda=50$] as functions of the evolution time $\lambda t$. The insets are zoomed-in plots of the probabilities in the middle duration of one period. Other parameters used are $\omega_{m}/\lambda=50.5$, $\omega_{e}/\lambda=250$, $g/\lambda=2.5$, $\xi=1.841$, and $\delta_{a}=-g_{a}/2$.}
\label{Fig6}
\end{figure}

When the electronic degree of freedom is detected on the bases $\vert \pm\rangle$, the cavity mode and the vibrational mode will collapse into the states
\begin{equation}
\vert\Psi_{\pm}(t)\rangle =\frac{1}{\sqrt{2P_{\pm}(t)}}\sum_{n,j=0}^{\infty}[A_{n,j}(t)\pm B_{n,j}(t)]\vert n\rangle _{a}\vert j\rangle_{b}\label{Psistate},
\end{equation}
where
\begin{equation}
P_{\pm}(t)=\frac{1}{2}\sum_{n,j=0}^{\infty}\vert[A_{n,j}(t)\pm B_{n,j}(t)]\vert ^{2}\label{proclose}
\end{equation}
are the detection probabilities corresponding to the states $\vert\Psi_{\pm}(t)\rangle$.

In Fig.~\ref{Fig6}, we show the time dependence of both the analytical measurement probabilities $\mathcal{P}_{\pm}(t)$ given by Eq.~(\ref{proapproximate}) and the measurement probabilities $P_{\pm}(t)$ defined in Eq.~(\ref{proclose}) at the selected vaule $\omega_{c}/\lambda=50$. Figure~\ref{Fig6} shows that the probabilities in the exact numerical case exhibit large magnitude oscillations in the two ends in one period, while the magnitude of oscillations decreases in the middle duration of one period. It can also be seen from Eq.~(\ref{proapproximate}) that, at the detection time $t_{s}$, the coherent amplitudes $\vert\alpha(t_{s})\vert$ and $\vert\beta(t_{s})\vert$ reach their peak values such that $\exp[-(\vert \alpha(t_{s})\vert^2+\vert \beta(t_{s})\vert^2)/2]\approx0$, then the probabilities $\mathcal{P}_{\pm}(t_{s})\approx1/2$. In addition, we see in Fig.~\ref{Fig6} that the numerical results (both the whole envelop and the details of the oscillation in the insets) are in good agreement with the analytical results for the used parameters.

\subsection{Fidelities between the approximate and exact states in the closed-system case \label{ssec:fidelityclose}}

\begin{figure}[tbp]
\center
\includegraphics[ width=0.47 \textwidth]{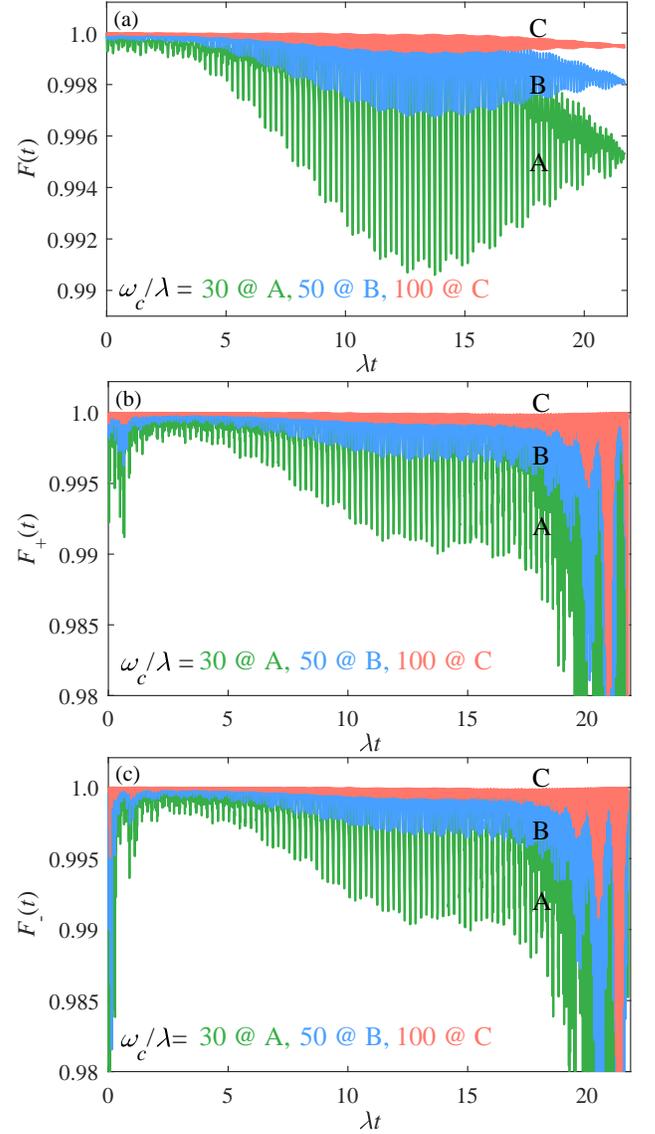}
\caption{(a-c) The fidelities $F(t)$ and $F_{\pm}(t)$ versus the evolution time $\lambda t$. From bottom to top, the curves correspond to the selected values of frequencies are $\omega_{c}/\lambda=30$ (green), $\omega_{c}/\lambda=50$ (blue), and $\omega_{c}/\lambda=100$ (red). Other parameters used in panels (a-c) are given by $\omega_{m}/\omega_{c}=1.01$, $\xi=1.841$, and $\delta_{a}=-g_{a}/2$.}
\label{Fig7}
\end{figure}

To evaluate the validity of the RWA made in the approximate Hamiltonian~(\ref{Hamiltrwa}), we calculate the fidelity $F(t)=\vert\langle\Psi(t)\vert\psi(t)\rangle\vert^{2}$ between the approximate state $\vert\psi(t)\rangle$ and the exact state $\vert\Psi(t)\rangle$, which are defined in Eqs.~(\ref{originalstatepsi}) and (\ref{Psinumexa}), respectively. The expression of this fidelity can be obtained as
\begin{equation}
F(t)=\frac{1}{2}\left\vert\sum_{n,j=0}^{\infty}e^{i\theta(t)-\frac{1}{2}(\vert\alpha(t)\vert^{2}+\vert\beta(t)\vert^{2})}A_{n,j}^{\ast}(t)\frac{\alpha^{n}(t)\beta^{j}(t)}{\sqrt{n!j!}}+B_{0,0}^{\ast }(t)\right\vert^{2}.
\end{equation}

Similarly, we can also evaluate the performance of the entangled-cat-state generation by calculating the fidelities between the generated states $\vert\Psi_{\pm}(t)\rangle$ (after the measurement) in Eq.~(\ref{Psistate}) and the target states $\vert \psi_{\pm}(t)\rangle$ (the entangled cat states) in Eq.~(\ref{psistate}). Using Eqs.~(\ref{Psistate}) and~(\ref{psistate}), the fidelities $F_{\pm}(t)=\vert\langle \Psi_{\pm}(t)\vert\psi_{\pm}(t)\rangle\vert^{2}$ can be obtained as
\begin{eqnarray}
F_{\pm }(t) &=&\frac{|\mathcal{M}_{\pm}(t)|^{2}}{P_{\pm}(t)}\left\vert
e^{i\theta(t)-\frac{1}{2}(\vert\alpha(t)\vert^{2})+\vert\beta(t)\vert^{2}}\sum_{n,j=0}^{\infty}[A_{n,j}^{\ast}(t)\pm B_{n,j}^{\ast}(t)] \right.\nonumber\\
&&\left. \times \frac{\alpha ^{n}(t)\beta ^{j}(t)}{\sqrt{2\times n!j!}}\pm\frac{1}{\sqrt{2}}[A_{0,0}^{\ast}(t)\pm B_{0,0}^{\ast}(t)]\right\vert^{2}.
\end{eqnarray}

\begin{figure}[tbp]
\center
\includegraphics[ width=0.47 \textwidth]{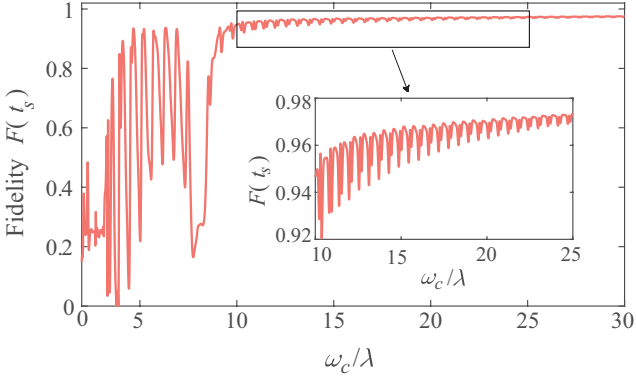}
\caption{The fidelity $F(t_{s})$ at the detection time $t_{s}$ versus the ratio $\omega_{c}/\lambda$. The inset shows that the fidelity could be larger than $0.95$ when $\omega_{c}/\lambda>20$. Other parameters used are given by $\omega_{m}/\omega_{c}=1.01$, $\omega_{e}/\lambda=250$, $g/\lambda=2.5$, $\xi=1.841$, and $\delta_{a}=-g_{a}/2$.}
\label{Fig8}
\end{figure}

In Fig.~\ref{Fig7}, we plot the fidelities $F(t)$ and $F_{\pm}(t)$ as functions of time $t$ when the ratio $\omega_{c}/\lambda$ takes various values. Here we can see that the fidelities exhibit fast oscillation because of the high-frequency oscillation terms $\exp[\pm i(\omega_{\pm}+ n\omega_{0})t]$. In particular, the oscillation frequency increases and the oscillation amplitude decreases with the increase of $\omega_{c}$. The lower envelop of the fidelity is larger for a larger value of $\omega_{c}/\lambda$, which means that the parameter condition of the RWA is satisfied when $\omega_{c}\gg\lambda$.

In Fig.~\ref{Fig8}, we plot the fidelity $F(t_{s})$ at the detection time $t_{s}$ as a function of the ratio $\omega_{c}/\lambda$. We can see that the numerical results agree better with the analytical results for a larger value of $\omega_{c}/\lambda$, which is consistent with the parameter conditions given in Eqs.~(\ref{paracondaa}) and~(\ref{paracondab}), under which the RWA is justified. We can also see from the inset of this figure that $F(t_{s})\geq0.95$ even for $\omega_{c}/\lambda\geq20$, which means that $\omega_{c}/\lambda\geq20$ is reasonable to guarantee the validity of RWA in our simulations.

\section{The open-system case \label{sec:opensys}}

In this section, we will study the generation of entangled cat states in the open-system case. In particular, the influence of the system dissipations on the fidelity and the success probability of the state generation, and the degree of entanglement of the generated states will be analyzed in detail.

\subsection{Quantum master equation and solution\label{sec:masterequa}}

In typical molecular cavity-QED systems, the energy scales of the cavity field and the molecule states are much larger than that of the bath temperatures, and hence the average thermal excitation numbers associated with these environments can be approximated as zero, which means that we can safely treat these environments of the system as vacuum baths. Within the Born-Markovian framework, the evolution of the system is governed by the quantum master equation
\begin{equation}
\dot{\rho}=-i[H(t),\rho]+\gamma_{e}\mathcal{D}[\sigma_{-}]\rho+\kappa\mathcal{D}[a]\rho+\gamma_{v}\mathcal{D}[b]\rho,\label{densmatgen}
\end{equation}
where the Hamiltonian $H(t)$ is given by Eq.~(\ref{Hamilt1}), the parameters $\kappa$, $\gamma _{v}$, and $\gamma _{e}$ are, respectively, the decay rates of the cavity field and the vibrational and electronic degrees of freedom of the molecule.
The superoperator $\mathcal{D}[o=\sigma_{-},a,b]\rho =o\rho o^{\dag}-(o^{\dag}o\rho+\rho o^{\dag}o)/2$ is the standard Lindblad dissipator that describes the dissipations of the cavity field and the molecule.

To solve quantum master equation~(\ref{densmatgen}), we expand the density matrix of the system in the Fock-state representation as
follows
\begin{eqnarray}
\rho(t) &=&\sum_{r,s=e,g}\sum_{m,k,n,j=0}^{\infty }\rho _{r,m,k,s,n,j}(t)|r\rangle_{p} |m\rangle _{a}|k\rangle _{b}\langle s|_{p}\langle n|_{a}\langle j|_{b}.\label{density matrix}
\end{eqnarray}
In this way, we can obtain the equations of motion for these density matrix elements, which can be solved numerically under the initial conditions. For our state-generation motivation, we consider the initial state $\vert+\rangle\vert 0\rangle_{a}\vert 0\rangle_{b}$, which corresponds to the initial value of these density matrix elements: $\rho_{e,m,k,e,n,j}(0)=\rho_{e,m,k,g,n,j}(0)=\rho_{g,m,k,e,n,j}(0)=\rho_{g,m,k,g,n,j}(0)=(1/2)\delta_{m,0}\delta_{k,0}\delta_{n,0}\delta_{j,0}$.
Based on the numerical solutions, the evolution of the density matrix of the system can be obtained, and then the properties of the generated states can be calculated.

\subsection{Fidelities between the approximate and exact states in the open-system case}

To evaluate the influence of the dissipation on the state generation, we calculate the fidelity between the obtained density matrix $\rho(t)$ in Eq.~(\ref{density matrix}) and the analytical state $\vert\psi(t)\rangle$ in Eq.~(\ref{totalpsistate}) as
\begin{eqnarray}
f(t)&=&\langle\psi(t)\vert\rho(t)\vert \psi(t)\rangle \nonumber \\
&=&\frac{1}{2}\sum_{m,k,n,j=0}^{\infty }\left[ \rho_{e,m,k,e,n,j}e^{-(\vert\beta(t)\vert^{2}+\vert\alpha(t)\vert^{2})}\Theta _{m,k,n,j}\right.   \nonumber \\
&&\left. +\rho _{e,m,k,g,n,j}e^{-i\theta (t)-\frac{1}{2}(\vert\beta(t)\vert^{2}+\vert\alpha(t)\vert^{2})}\Theta_{m,k,n,j}\delta _{n,0}\delta _{j,0}\right.   \nonumber \\
&&\left. +\rho _{g,m,k,e,n,j}e^{i\theta(t)-\frac{1}{2}(\vert\beta(t)\vert^{2}+\vert\alpha(t)\vert^{2})}\Theta_{m,k,n,j}\delta _{m,0}\delta _{k,0}\right.   \nonumber \\
&&\left. +\rho _{g,m,k,g,n,j}\Theta _{m,k,n,j}\delta _{m,0}\delta_{k,0}\delta _{n,0}\delta _{j,0}\right],
\end{eqnarray}
where $\Theta_{m,k,n,j}=[\alpha^{\ast}(t)]^{m}[\beta^{\ast}(t)]^{k}\alpha^{n}(t)\beta^{j}(t)/\sqrt{(k!m!j!n!)}$ is introduced.

\begin{figure}[tbp]
\center
\includegraphics[ width=0.47 \textwidth]{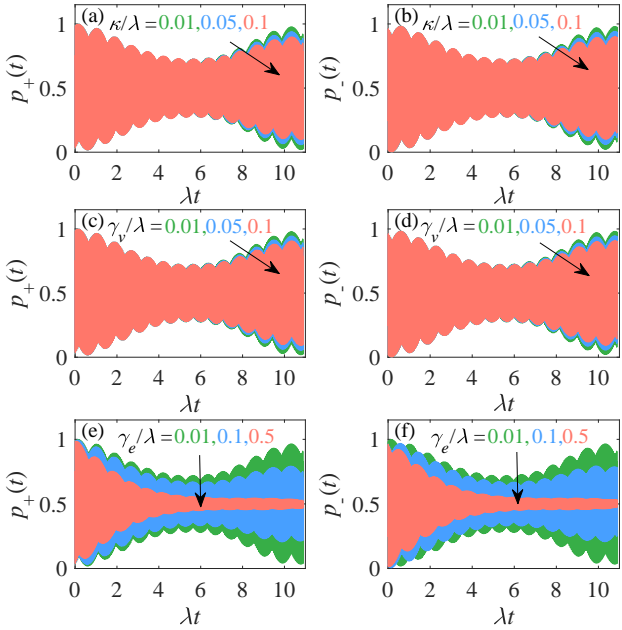}
\caption{The probabilities $p_{\pm}(t)$ as functions of the evolution time $\lambda t$ in various cases. (a), (b) $\gamma_{v}/\lambda=\gamma_{e}/\lambda=0.001$ and $\kappa/\lambda=0.01$, $0.05$, and $0.1$. (c), (d) $\kappa/\lambda=\gamma_{e}/\lambda=0.001$ and $\gamma_{v}/\lambda=0.01$, $0.05$, and $0.1$. (e), (f) $\kappa/\lambda=\gamma_{v}/\lambda=0.001$ and $\gamma_{e}/\lambda=0.01$, $0.1$, and $0.5$. Other parameters used are $\omega_{c}/\lambda=100$, $\omega_{m}/\lambda=101$, $\omega_{e}/\lambda=250$, $g/\lambda=2.5$, $\xi=1.841$, and $\delta_{a}=-g_{a}$.}
\label{Fig9}
\end{figure}

For generation of entangled cat states, the projective measurement on the electronic states $\vert\pm\rangle$ should be performed, and then the corresponding density matrices of the cavity field and the vibrational mode become
\begin{equation}
\rho^{(\pm)}(t)=\frac{1}{2p_{\pm }(t)}\sum_{m,k,n,j=0}^{\infty}\Lambda_{m,k,n,j}^{(\pm)}(t)\vert m\rangle_{a}\vert k\rangle_{b}\langle n\vert_{a}\langle j\vert_{b},\label{matrixpm}
\end{equation}
where we introduce the variables
\begin{eqnarray}
\Lambda _{m,k,n,j}^{(\pm )}(t) &=&\left[ \rho _{e,m,k,e,n,j}(t)+\rho_{g,m,k,g,n,j}(t)\pm \rho _{e,m,k,g,n,j}(t)\right.  \nonumber\\
&&\left. \pm \rho _{g,m,k,e,n,j}(t)\right]\Theta_{m,k,n,j}
\end{eqnarray}
and the probabilities for detecting the electronic states $\vert\pm\rangle$,
\begin{equation}
p_{\pm}(t)=\frac{1}{2}\sum_{m,k=0}^{\infty}\Lambda_{m,k,n,j}^{(\pm)}(t).\label{proopen}
\end{equation}
Accordingly, the fidelities between the generated states $\rho^{(\pm)}(t)$ and the target states $\vert\psi_{\pm}(t)\rangle$ can be calculated as
\begin{eqnarray}
f_{\pm}(t)&=&\langle\psi_{\pm}(t)\vert\rho^{(\pm)}(t)\vert \psi_{\pm}(t)\rangle \nonumber\\
&=&\frac{\vert \mathcal{M}_{\pm}(t)\vert^{2}}{2p_{\pm}(t)}\sum_{m,k,n,j=0}^{\infty }\left[
e^{-(\vert\beta(t)\vert^{2}+\vert\alpha(t)\vert^{2})}\Lambda_{m,k,n,j}^{(\pm)}(t)\right.\nonumber\\
&&\left.\pm e^{-i\theta(t)}e^{-\frac{1}{2}(\vert\beta(t)\vert^{2}+\vert\alpha(t)\vert^{2})}\Lambda_{m,k,n,j}^{(\pm)}(t)\delta_{0,n}\delta_{0,j}\right.\nonumber\\
&&\left.\pm e^{i\theta(t)}e^{-\frac{1}{2}(\vert\beta(t)\vert^{2}+\vert\alpha(t)\vert^{2})}\Lambda_{m,k,n,j}^{(\pm)}(t)\delta_{0,m}\delta_{0,k}\right.\nonumber\\
&&\left.+\Lambda_{m,k,n,j}^{(\pm)}(t)\delta_{0,m}\delta_{0,n}\delta_{0,k}\delta_{0,j}\right].
\end{eqnarray}

\begin{figure}[tbp]
\center
\includegraphics[bb=0 0 579 450, width=0.47  \textwidth]{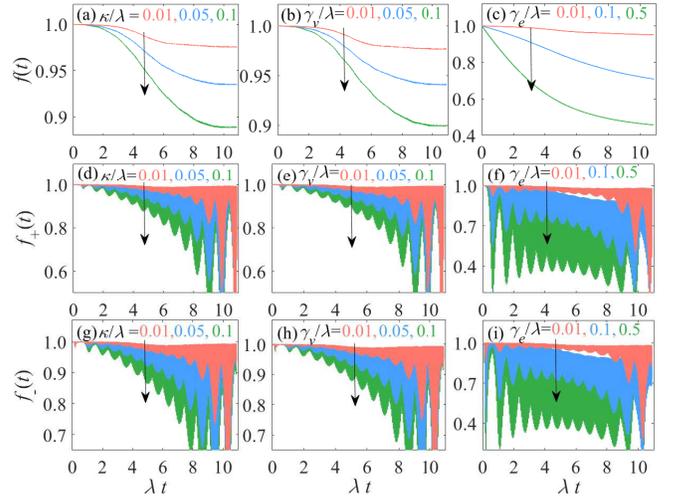}
\caption{The fidelities $f(t)$ and $f_{\pm}(t)$ as functions of the evolution time $\lambda t$ in various cases. (a), (d), (g) $\gamma_{v}/\lambda=\gamma_{e}/\lambda=0.001$ and $\kappa/\lambda=0.01$, $0.05$, and $0.1$. (b), (e), (h) $\kappa/\lambda=\gamma_{e}/\lambda=0.001$ and $\gamma_{v}/\lambda=0.01$, $0.05$, and $0.1$. (c), (f), (i) $\kappa/\lambda=\gamma_{v}/\lambda=0.001$ and $\gamma_{e}/\lambda=0.01$, $0.1$, and $0.5$. Other parameters used are the same as those in Fig.~\ref{Fig9}.}
\label{Fig10}
\end{figure}

To analyze the state generation probabilities in the open system case, in Fig.~\ref{Fig9}, we display the time dependence of the probabilities $p_{\pm}(t)$ at selected values of the scaled cavity-field decay rate $\kappa/\lambda$, the scaled molecular vibration dissipation rate $\gamma_{v}/\lambda$, and the scaled electronic transition dissipation rate $\gamma_{e}/\lambda$. Figures~\ref{Fig9}(a)-\ref{Fig9}(d) show that the probabilities $p_{\pm}(t)$ oscillate rapidly, which is mainly caused by the free evolution of the electronic state, as shown by the phase factor $\theta(t)$. With the evolution of the system, the amplitude of the oscillation envelop decreases gradually. In the intermediate duration of $\lambda t\approx4.5-6.5$, the value of the probabilities $p_{+}(t)\approx p_{-}(t)\rightarrow1/2$. The amplitude of the oscillation envelop will revive in the latter part of a period. However, the amplitude of the revival envelop is small for a large value of the dissipation rates. This phenomenon can be seen more clearly in Figs.~\ref{Fig9}(e)-\ref{Fig9}(f). Here we see that the greater the dissipation rate is, the smaller amplitude of the oscillation envelop for the probabilities is. When the electronic state dissipation rate $\gamma_{e}/\lambda=0.5$, the values of the probabilities $p_{\pm}(t)$ approach to a stable value around $1/2$ and the revival phenomenon disappears.

In Fig.~\ref{Fig10}, the fidelities $f(t)$ and $f_{\pm}(t)$ are plotted as functions of evolution time $\lambda t$ with different values of the scaled decay rates $\kappa/\lambda$, $\gamma_{v}/\lambda$, and $\gamma_{e}/\lambda$. As shown in Figs.~\ref{Fig10}(a)-\ref{Fig10}(c), the fidelities experience decay dynamics with the evolution of the system. The fidelity $f(t)$ decays faster for lager values of these decay rates. In Figs.~\ref{Fig10}(d)-\ref{Fig10}(i), we show the time dependence of $f_{\pm}(t)$ at various values of the decay rates $\kappa/\lambda$, $\gamma_{v}/\lambda$, and $\gamma_{e}/\lambda$. Here we can see that the fidelities exhibit fast oscillation and that the values of $f_{\pm}(t)$ depend on the dissipation rates $\kappa/\lambda$, $\gamma_{v}/\lambda$, and $\gamma_{e}/\lambda$, similar to the behavior of the fidelity $f(t)$. The lower envelope of these fidelities is smaller for lager values of these decay rates. Moreover, the fidelities $f_{\pm}(t)$ experience a sudden decrease at the end of a period, similar to the closed-system case.

\begin{figure}[tbp]
\center
\includegraphics[ width=0.47 \textwidth]{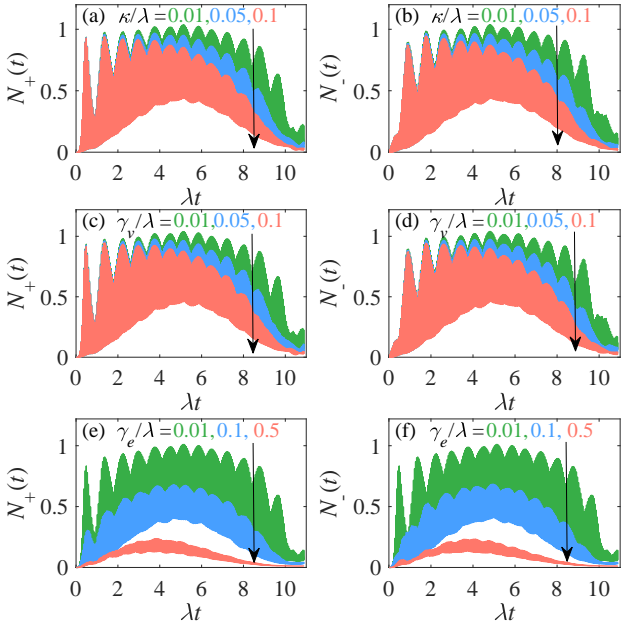}
\caption{The logarithmic negativity $N_{\pm}(t)$ as functions of the scaled evolution time $\lambda t$ in various cases. (a), (b) $\gamma_{v}/\lambda=\gamma_{e}/\lambda=0.001$ and $\kappa/\lambda=0.01$, $0.05$, and $0.1$. (c), (d) $\kappa/\lambda=\gamma_{e}/\lambda=0.001$ and $\gamma_{v}/\lambda=0.01$, $0.05$, and $0.1$. (e), (f) $\kappa/\lambda=\gamma_{v}/\lambda=0.001$ and $\gamma_{e}/\lambda=0.01$, $0.1$, and $0.5$. Other parameters used are the same as those in Fig.~\ref{Fig9}.}
\label{Fig11}
\end{figure}

\subsection{Entanglement between the cavity-field mode and the molecular vibrational mode}

In the open-system case, the logarithmic negativity is used for quantizing the quantum entanglement between the cavity mode and the molecular vibration for the generated density matrices $\rho^{(\pm)}(t)$. In terms of Eqs.~(\ref{lognegtivity}),~(\ref{densmatgen}), and~(\ref{matrixpm}), the logarithmic negativity of the states $\rho^{(\pm)}(t)$ can be solved numerically. In Fig.~\ref{Fig11}, we plot the logarithmic negativity $N_{\pm}(t)$ as functions of the evolution time $\lambda t$ when the dissipation rates of the system take different values. Here we can see that the logarithmic negativity oscillates very fast with the evolution of the system. Similar to the pure-state case in Fig.~\ref{Fig3}, around the decoupling times $t=2\pi/\delta_{a}$, the cavity field and the vibrational mode decouple and the logarithmic negativity $N_{\pm}(t)$ become zero. At the initial time of each cycle, the logarithmic negativity first increases rapidly with the increase of time $\lambda t$ and reaches the maximum in the intermediate duration of $\lambda t\approx 3-7$, then the logarithmic negativity begins to decrease smoothly until $0$. For larger decay rates, the maximum value which the logarithmic negativity can reach is smaller and the logarithmic negativity decays faster.

\section{Discussions on the experimental condition and challenge \label{sec:discussion}}

In this section, we present some discussions on the experimental feasibiility and challenge for implementation of this state-generation scheme with physical systems. In principle, the physical mechanism in this work is general and hence the method for generation of macroscopic entangled cat states can be implemented in various physical systems which possess the same interactions described by the system Hamiltonians. Below we focus our discussions on the molecular cavity-QED systems, in which both the coupling between the electronic states and the vibrational states and the interaction between the cavity field and the molecular vibrational mode can be realized in current experimental conditions~\cite{Benz2016,Chikkaraddy2016}. The realistic parameters of this system could be taken as: $\omega_{c}\approx\omega_{m}\sim2\pi\times 10$-$52$ THz, $\omega_{e}\approx\pi\times 125$-$250$ THz, $g\approx 2\pi\times0.4$ - $2$ THz, $\lambda\approx 2\pi\times0.1$-$1$ THz, and $\kappa_{r}\approx\gamma_{e}\approx\gamma_{M}\sim2\pi\times1$-$100$ GHz. In these systems, the energy scales of these degrees of freedom are much larger than the enenrgy scale corresponding to the room temperature, and hence the environments of these subsystems can be considered as vacuum baths. In our simulations, we used the scaled parameters: $g/\lambda= 2.5$, $\omega_{c}/\lambda= 30$-$100$, and $\kappa_{r}/\lambda\approx\gamma_{e}/\lambda\approx\gamma_{M}/\lambda=0.001$-$0.1$, which are in consistent with the above parameter analyses. In addition, the resonance frequency modulations of the cavity mode and the molecular vibration are introduced in this work to enhance the conditional displacement. Therefore, we should analyze the experimental realization of the frequency modulations. The modulation parameters $\omega_{0}$ and $\xi$ should be chosen to satisfy the RWA condition and make sure the values of $\xi$, $n_{a}$, and $n_{b}$ to be consistent. In our simulation, we choose $n_{a}=1$, $n_{b}=1$, and $\xi=1.841$. Note that smaller values of the modulation frequency $\omega_{0}$ can be chosen by choosing larger values of the Floquet-sideband indexes $n_{a}$ and $n_{b}$. It has been analyzed that the realization of the cavity-frequency modulations is accessible with current or near-future experimental conditions~\cite{Liao2016b}. However, the realization of the resonance frequency modulation for the molecular vibration remains a challenge. Therefore, proper experimental techniques for modulating the molecular vibration should be developed to enrich the quantum manipulation means in molecular cavity-QED systems.

\section{CONCLUSIONS \label{sec:conclusion}}

In conclusion, we have proposed a scheme to generate entangled cat states in the molecular cavity-QED system, which is composed of an organic molecule coupled to a single-mode cavity field. We have shown that the amplitude of the entangled coherent-state components can be largely enhanced by introducing monochromatic frequency modulations to the the cavity field and the molecule vibration. The Floquet-sideband modes induced by these frequency modulations assist the physical displacement mechanism and effectively enhance the near-resonant displacement of the cavity field and the molecular vibration. As a result, the macroscopic cat states can be generated in the sense that the displacement amplitudes could be larger than the corresponding zero-point fluctuations. We have checked the validity of the approximate Hamiltonian by evaluating the fidelity between the approximate and exact states. We have also studied the quantum properties of the generated states by calculating the joint Wigner function and the quantum entanglement. The influence of the system dissipations on the fidelity and entanglement have been analyzed with the quantum master equation method. Some discussions on the experimental implementation of this scheme with current experimental conditions have been presented.

\begin{acknowledgments}
J.-Q.L. is supported in part by National Natural Science Foundation of China (Grants No.~11822501, No.~11774087, and No.~11935006), Natural Science Foundation of Hunan Province, China (Grant No.~2017JJ1021), and Hunan Science and Technology Plan Project (Grant No.~2017XK2018). J.-F.H. is supported in part by the National Natural Science Foundation of China (Grant No.~11505055 ) and Scientific Research Fund of Hunan Provincial Education Department (Grant No. 18A007). 
\end{acknowledgments}

\appendix*
\section{Derivation of Eq.~(\ref{unitary})}

In this appendix, we present a detailed derivation of the unitary evolution operator $U_{e}(t) $ given in Eq.~(\ref{unitary}). For the Hamiltonian $H_{\textrm{RWA}}(t)=H_{e}(t)\otimes \vert e\rangle \langle e\vert +H_{g}\otimes \vert g\rangle \langle g\vert $, the unitary evolution operator $U_{\textrm{RWA}}(t)$ is governed by the equation of motion $i\partial U_{\textrm{RWA}}(t)/\partial t=H_{I}(t)U_{\textrm{RWA}}(t)$ subject to the initial condition $U_{\textrm{RWA}}(0)=I$, where $I$ is the identity matrix in the Hilbert space of the system. Formally, the expression of the evolution operator $U_{\textrm{RWA}}(t)$ can be written as
\begin{eqnarray}
U_{\textrm{RWA}}(t) &=&\mathcal{T}\exp \left\{\int_{0}^{t}[-iH_{\textrm{RWA}}(\tau )]d\tau\right\} \nonumber\\
&=&\mathcal{T}\exp\left\{-i\int_{0}^{t}[H_{e}(\tau) \otimes\vert e\rangle \langle e\vert +H_{g}(\tau) \otimes \vert g\rangle\langle g\vert] d\tau\right\}  \nonumber\\
&=&U_{e}(t)\otimes\vert e\rangle\langle e\vert +U_{g}\otimes \vert g\rangle\langle g\vert,
\end{eqnarray}
where $U_{e}(t) =\mathcal{T}\exp[-i\int_{0}^{t}H_{e}(\tau)d\tau]$ and $U_{g}=I$ are, respectively, the time evolution operators of the cavity-field and vibration modes when the electronic degree of freedom is in the excited state $\vert e\rangle$ and ground state $\vert g\rangle$, with ``$\mathcal{T}$" being the time-ordering operator.

According to the Magnus expansion, the operator $U_{e}(t)$, which is determined by the equation of motion $i\partial U_{e}(t)/\partial t=H_{e}(t)U_{e}(t)$, can be expressed as
\begin{equation}
U_{e}(t)=\exp[\Lambda(t)],\hspace{1 cm}  \Lambda(t)=\sum_{k=1}^{\infty }\Lambda_{k}(t),
\end{equation}
with
\begin{eqnarray}
\Lambda_{1}(t)&=&\int_{0}^{t}[-iH_{e}(t_{1})]dt_{1},\nonumber\\
\Lambda_{2}(t)&=&\frac{1}{2}\int_{0}^{t}dt_{1}\int_{0}^{t_{1}}dt_{2}[-iH_{e}(t_{1}),-iH_{e}(t_{2})],\nonumber\\
\Lambda_{3}(t)&=&\frac{1}{6}\int_{0}^{t}dt_{1}\int_{0}^{t_{1}}dt_{2}\int_{0}^{t_{2}}dt_{3}\nonumber\\
&&\times\left([-iH_{e}(t_{1}),[-iH_{e}(t_{2}),-iH_{e}(t_{3})]]\right.\nonumber\\
&&\left.+[-iH_{e}(t_{3}),[-iH_{e}(t_{2}),-iH_{e}(t_{1})]]\right),\nonumber\\
\Lambda_{k>3}(t)&=&\cdots.
\end{eqnarray}
In the present case, the value of the commutator $[H_{e}(t_{1}),H_{e}(t_{2})]$ is c-number, and then it is sufficient to keep the Magnus expansion up to the second order because of $\Lambda_{k>2}(t)=0$. As a result, the unitary evolution operator associated with the Hamiltonian $H_{e}(t)$ can be expressed as
\begin{equation}
U_{e}(t)=\exp[\Lambda(t)]=\exp[\Lambda_{1}(t)]\exp[\Lambda_{2}(t)],
\end{equation}
with
\begin{subequations}
\begin{align}
\Lambda _{1}(t)=&\eta^{\ast}(t)a-\eta(t)a^{\dag}+\zeta(t)b^{\dag}-\zeta^{\ast }(t)b,\\
\Lambda_{2}(t)=&\theta_{a}(t)+\theta_{b}(t),
\end{align}
\end{subequations}
where $\theta_{a}(t)=g_{a}^{2}[\delta_{a}t-\sin(\delta_{a}t)]/\delta_{a}^{2}$ and $\theta_{b}(t)=g_{b}^{2}[\delta_{b}t-\sin(\delta_{b}t)]/\delta_{b}^{2}$ are the phase factors of cavity-field mode and molecular vibration, respectively. The variables $\eta(t)=g_{a}(1-e^{i\delta_{a}t})/\delta_{a}$ and $\zeta(t)=g_{b}(1-e^{i\delta_{b}t})/\delta_{b}$ are, respectively, the displacement amplitudes of the cavity-field mode and the molecular vibration.

\end{document}